# 3D Brownian Diffusion of Submicron-sized Particle Clusters

*Martin Hoffmann,[‡] Claudia S. Wagner,[‡] Ludger Harnau,[§,†] and Alexander Wittemann[‡,]* *

‡ *Physikalische Chemie I, Universität Bayreuth, Universitätsstr. 30, 95440 Bayreuth, Germany,*

§ *Max-Planck-Institut für Metallforschung, Heisenbergstr. 3, D-70569, Stuttgart, Germany,*

† *Institut für Theoretische und Angewandte Physik, Universität Stuttgart, Pfaffenwaldring 57, D-70569, Stuttgart, Germany.*

Alexander.Wittemann@uni-bayreuth.de

**RECEIVED DATE**

CORRESPONDING AUTHOR FOOTNOTE

E-mail: Alexander.Wittemann@uni-bayreuth.de. Telephone: +49 921 55 2776. Fax: +49 921 55 2780.

**ABSTRACT** We report on the translation and rotation of particle clusters made through the combination of spherical building blocks. These clusters present ideal model systems to study the motion of objects with complex shape. Since they could be separated into fractions of well-defined configurations on a sufficient scale and because their overall dimensions were below 300 nm, the translational and rotational diffusion coefficients of particle doublets, triplets and tetrahedrons could be determined by a combination of polarized dynamic light scattering (DLS) and depolarized dynamic light scattering (DDLS). The use of colloidal clusters for DDLS experiments overcomes the limitation of earlier experiments on the diffusion of complex objects near surfaces because the true 3D diffusion can be studied. Knowing the exact geometry of the complex assemblies, different hydrodynamic models for calculating the diffusion coefficients for objects with complex shapes could be applied. Because hydrodynamic friction must be restricted to the cluster surface the so-called shell model, in which the surface is represented as a shell of small friction elements, was most suitable to describe the dynamics. A quantitative comparison of the predictions from theoretical modeling with the results obtained by DDLS showed an excellent agreement between experiment and theory.

**INTRODUCTION** Translational and rotational diffusion of colloidal particles was extensively studied by experiment,[1-7] simulation[8] and theory[9-13] over the past decades. In most of these studies, colloidal particles with simple shapes such as spheres,[14-18] ellipsoids,[19,20] rods,[21-26] and platelets[27] were used. The hydrodynamic properties of well-defined model colloids with shapes that differ from these simple geometries are important to understand the diffusion of objects with complex shapes.[28] The dynamics of complex particles is fundamental to the understanding of many practical problems such as biodistribution, sedimentation, coagulation, floatation and rheology.[29] Hence, Granick and coworkers prepared different planar clusters from micron-sized particles.[30,31] Such particle clusters are ideal candidates for the study of the motion of complex objects because they exhibit well-defined geometries.[32] The 2D diffusion of the clusters could be studied by video microscopy because the overall dimensions of the clusters were in the micron range.[30] The planar particle assemblies were prepared through evaporation of a suspension of silica microspheres on a microscope slide. The randomly distributed planar aggregates were cemented together and resuspended in an aqueous solution.[30] This technique is limited to micron-sized planar assemblies and small quantities of clusters.

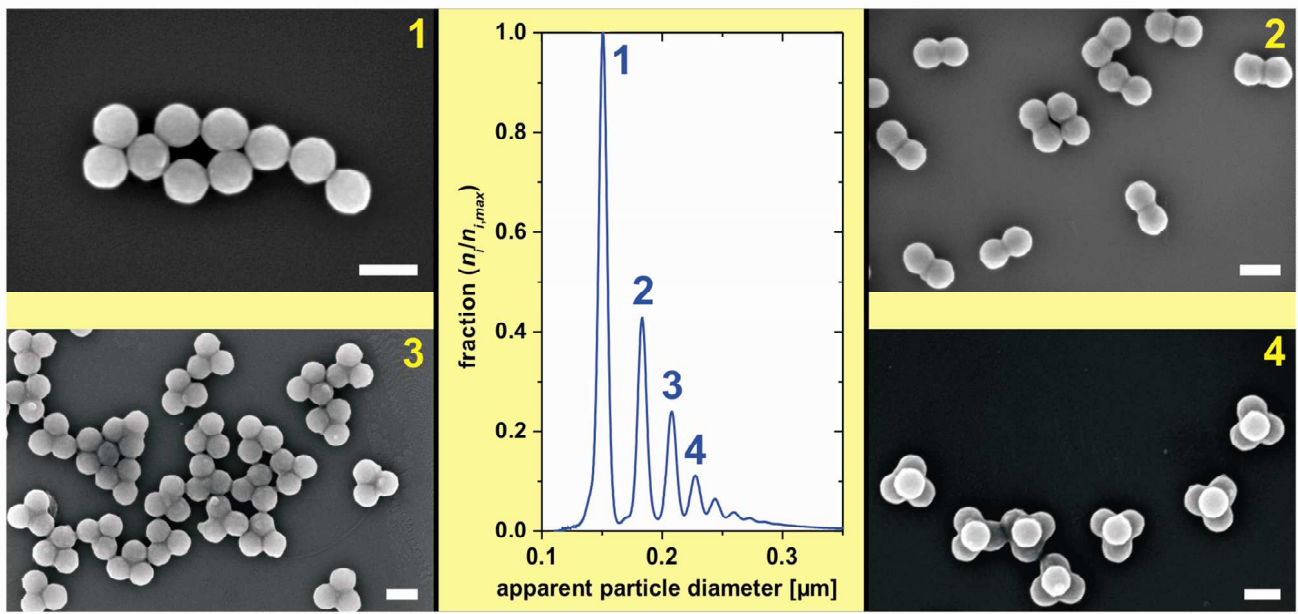

**Figure 1.** Sample preparation: The clusters were fabricated through the aggregation of particles adsorbed onto narrow-dispersed emulsion droplets. The statistical distribution of the building blocks onto the droplets led to clusters of different size. The number-weighted size distribution of the clusters obtained by analytical disk centrifugation shows that essentially large amounts of small clusters were formed (center). Because of their different sedimentation velocity, the clusters could be separated through centrifugation in a density gradient into fractions of single particles (1), particle doublets (2), triplets (3), and tetrahedrons (4). The FESEM micrographs demonstrate that suspensions of uniform clusters were obtained which present model systems for particles with complex shapes. Scale bars are 200 nm.

Velev and coworkers prepared 3D assemblies of microspheres using emulsion droplets as a template for the cluster formation.[33,34] Pine, Bibette and coworkers developed this technique further for the production of clusters having well-defined configurations.[35,36] Packing of the microspheres was accomplished through the agglomeration of the particles adsorbed onto the surface of macroemulsion droplets.[37] Recently, we combined this approach with basic principles established in the miniemulsion technique.[38] Colloidal clusters with overall dimensions below 300 nm could be obtained using narrow-dispersed emulsion droplets with diameters of 1.9 µm prepared through power ultrasonication and monodisperse spherical building blocks with diameters of 154 nm.[38] Clusters of these dimensions underlie Brownian motion which prevails over gravitational forces. The dynamics of these submicron-sized clusters with well-defined configurations can be thus studied by dynamic light scattering techniques. Hence, they present excellent model systems to study the diffusion of particles with complex shapes.

Polarized dynamic light scattering (DLS) became a routine technique to measure translational diffusion coefficients of submicron-sized particles. In DLS, the incident light is usually vertically polarized. The scattered light is dominated by the vertically polarized contribution but it can contain a horizontally polarized contribution as well.[22,39] In depolarized dynamic light scattering (DDLS) experiments, the latter contribution to the scattered light is measured through a horizontally oriented polarizer, such as a Glan-Thompson prism.[39] Pecora and coworkers could demonstrate that both translational and rotational diffusion coefficients can be derived from the contribution of the horizontally polarized scattered light.[22]

DDLS has been frequently applied to small molecules[40-42] but less often to large molecules or particles because of their relatively weak depolarized signal.[19,26] However, different theoretical models and computational procedures have been proposed for the calculation of the hydrodynamic properties of complex particles. A comprehensive overview of model building and hydrodynamic calculation is given in Ref. [28,43,44].

Here we present for the first time a study of both the translational and rotational diffusion of submicron-sized colloidal clusters consisting of up to four building blocks through a combination of DLS and DDLS. This technique could be used because the clusters underlie Brownian motion. Hence, in contrast to self diffusion measurements of micron-sized objects by microscopy,[30,45] the true 3D diffusion of complex colloids excluding wall effects can be studied by DDLS. Moreover, different models[28,44] that have been proposed for the description of the hydrodynamic properties of complex particles were probed to predict the translational and rotational diffusion coefficients of the clusters with regard to their configuration. The theoretical results were then compared with the experimental results to get a clear understanding of the hydrodynamics of the complex assemblies.

**RESULTS AND DISCUSSION**.

**Particle clusters**

We studied dilute aqueous suspensions of particle clusters with a specific number of constituents $N$ such as particle doublets ($N$ = 2), triplets ($N$ = 3) and tetrahedrons ($N$ = 4). These clusters consisted of amino-modified polystyrene particles of 154 in diameter (Figure 1). The assembly into clusters was accomplished using narrow-dispersed emulsion droplets as templates.[38] The colloidal building blocks were adsorbed onto the oil droplets because the adsorption lowers the interfacial energy due to the Pickering effect. Subsequent evaporation of the oil causes capillary forces which make the particles to pack together. This process leads to clusters of well-defined configurations (Figure 1) which are believed to result from a preorientation of the particles already at the droplet surface.[37,46]

Figure 1 shows that the suspension of the clusters can be separated by centrifugation into fractions of uniform clusters because of their different sedimentation velocity in a density gradient.[38] The analysis of the dynamics was restricted to clusters made from up to four constituents because these clusters have only one distinct configuration. Cluster consisting of more than four particles may have more than one configuration. For example, five spherical building blocks can be assembled into triangular dipyramids or square pyramids. Moreover, the smaller clusters could be prepared in scales which are sufficient for scattering experiments, and they can be separated into uniform fractions by centrifugation because of the large difference in mass of small assemblies (Figure 1).

**DDLS experiments**

In the following, we discuss the DLS and DDLS analysis of the submicron-sized clusters, which were used to determine their translational and rotational diffusion coefficients $D^T$ and $D^R$, respectively. The principle of the experimental setup is shown in Figure 2A. In both DLS and DDLS, the incident light was vertically polarized. The DDLS intensity autocorrelation functions (Figure 2B) presented the sum of two discrete exponential decays, where the slow mode was related to the translational diffusion while the fast mode originated from translational and rotational diffusion. The relaxation rates of the slow and the fast modes, $\Gamma_{slow}$ and $\Gamma_{fast}$, of the autocorrelation functions can be expressed as follows:[4]

$$\Gamma_{slow} = D^T q^2 \qquad (1)$$

$$\Gamma_{fast} = 6 D^R + D^T q^2 \qquad (2)$$

In DLS, the autocorrelation functions are dominated by the slow mode with the relaxation rate $\Gamma_{slow}$ (equation (1)). The contribution of the fast mode to the autocorrelation function is poor and can be neglected as long as the scattering objects have dimensions which are of the same order than the inverse absolute value of the scattering vector $1/q$ ($q = 4\pi n/\lambda \sin(\theta/2)$, where $n$ is the refractive index of the solvent, $\lambda$ the wavelength, and $\theta$ the scattering angle). This is the case for the clusters with dimensions below 300 nm. Hence, DLS could be used to study the translational diffusion of the clusters but not for the analysis of their rotational dynamics. Therefore we used DDLS to study the rotation of the clusters as well.

In DDLS, the horizontally polarized component of the light scattered by the clusters is detected through a polarizer (vH detection).[39] This contribution to the scattered light is much smaller than the intensity of the vertically polarized scattered light (vV detection) measured in a DLS experiment. This did not present an obstacle for the study of the cluster hydrodynamics because the concentrations of the cluster suspensions were still in the dilute regime but high enough to record intensity autocorrelation functions in vH detection (Figure 2B). In principle, the depolarized autocorrelation functions of the clusters should be characterized by a single exponential decay with the relaxation rate $\Gamma_{fast}$ following equation (2). However, the autocorrelation functions were the sum of two discrete experimental decays. This was further corroborated by CONTIN analysis[47] of the autocorrelation functions which was used to calculate the distribution of the relaxation times $A(\tau)$. In all cases, bimodal distributions of the relaxation times $\tau$ were obtained (Figure 2C). The fast mode originated from translational and rotational diffusion of the clusters according to equation (2) (Figure 2D, see also Supporting Information Figures 1 and 2)). Hence, it can be used to obtain $D^T$ and $D^R$. The second mode, *i.e.* the slow mode, emerges from vertically polarized scattered light as a consequence of the limited extinction ratio of the Glan-Thomson polarizer ($10^{-5}$). Because this mode follows equation (1) it provides an additional access to $D^T$ (Figure 2D).

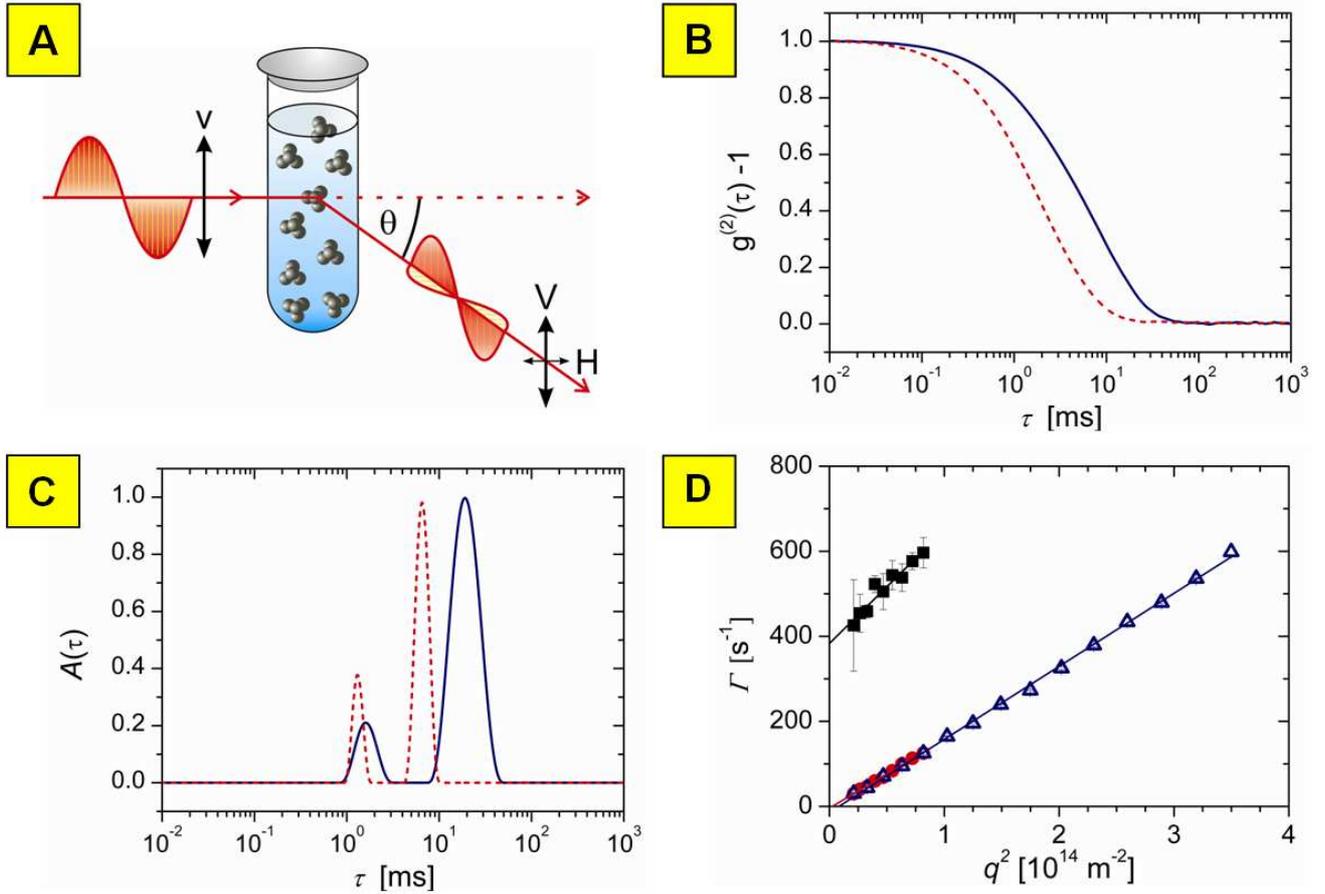

**Figure 2.** Depolarized dynamic light scattering (DDLS) experiments of particle tetrahedrons ($N = 4$): (A) Schematic representation of the experiment: The incident light is vertically (v) polarized. Moreover, the signal of the light scattered by the clusters is mainly governed by vertically polarized light (V) as well. In DDLS, a horizontally oriented polarizer is used to measure the horizontally polarized component (H) of the scattered light. The latter is zero for isotropic particles but nonzero for scatters with optical or shape anisotropy. (B) Depolarized intensity autocorrelation functions ($g^{(2)}(\tau) - 1$) measured at 30° (blue line) and 40° (red dashed line) for a suspension of particle tetrahedrons. (C)

Relaxation time distributions $A(\tau)$ (CONTIN Plots) as derived from the autocorrelation functions. The right peak (slow relaxation mode) originates from the translational diffusion of the clusters, whereas the left peak (fast relaxation mode) contains information on both the translation and rotation of the complex assemblies. (D) Relaxation rates $\Gamma$ as the function of the square of the absolute value of the scattering vector $q$: Slow modes of DLS (blue triangles) and DDLS (red spheres); fast mode of DDLS (black spheres). The linear correlations follow equation (1) (slow modes) and equation (2) (fast modes) respectively, which in turn give access to the translational and rotational diffusion coefficients.

The particle clusters are bearing amino groups on their surface which leads to electrostatic stabilization. Electrostatic repulsion among the clusters might affect the dynamics.[48,49] Hence, we immersed the clusters in solutions of 10 mM NaCl to screen the charges. The results did not differ from the values of $D^T$ and $D^R$ measured in pure water. Hence, all further experiments could be performed in water because the electrostatics did not influence the dynamics of the clusters in the dilute regime.

The building blocks of the clusters should be optically isotropic because of their spherical shape. However, even for these particles a depolarized signal could be measured. Despite the low polydispersity of the building blocks of the clusters, there might be slight deviations from either the spherical shape or an uneven distribution of the amino groups on the surface, which makes the particles optically anisotropic. The depolarized signal caused by such effects is rather poor. At a scattering angle of 40°, the contribution of the fast mode to the intensity of the scattered light is only 3 %. Nonetheless, it could be used to determine $D^R$ in addition to $D^T$ (see Supporting Information Figure 1). Of course, $D^R$ is affected by a larger error in this case than for the particle doublets and triplets which have an anisotropic shape. The shape anisotropy of the doublets and triplets result in large depolarized signals, *i.e.* 30 % and 21 %, respectively of the total intensity of the scattered light. The intensity of the depolarized signal increases with the shape anisotropy of the assemblies (see Supporting Information Figure 3). Therefore DDLS is a suitable method to especially study the dynamics of dumbbell-shaped [3] or rod-like particles.[26]

According to equations (1) and (2) the linear regressions of $\Gamma_{fast}$ and $\Gamma_{slow}$ on $q^2$ shown in Figure 2D should have the same slopes, *i.e.* the translational diffusion coefficient $D^T$. This holds as long as the translation and rotation motions of the clusters are decoupled. We observed a perfect agreement of $D^T$ as obtained from DLS and from both modes for the particle doublets and triplets, which in turn confirms the decoupling of rotational from translational diffusion (see Supporting Information Figure 2). For the tetrahedrons, the slopes of the linear regressions shown in Figure 2D slightly differ. Because of the low volume fractions of the cluster suspensions of less than $10^{-5}$, interparticular interactions and a coupling between rotation and diffusion can be excluded.[50] The deviation might be due to the larger experimental error of the DDLS measurement of tetrahedrons because a tetrahedron has rather small shape anisotropy as compared to a doublet and a triplet. However, the building blocks are slightly optically anisotropic. Moreover, they may also slightly differ in size which contributes to the shape anisotropy. For tetrahedrons, the fast mode contributes 10 % to the total scattering intensity of the DDLS experiment.

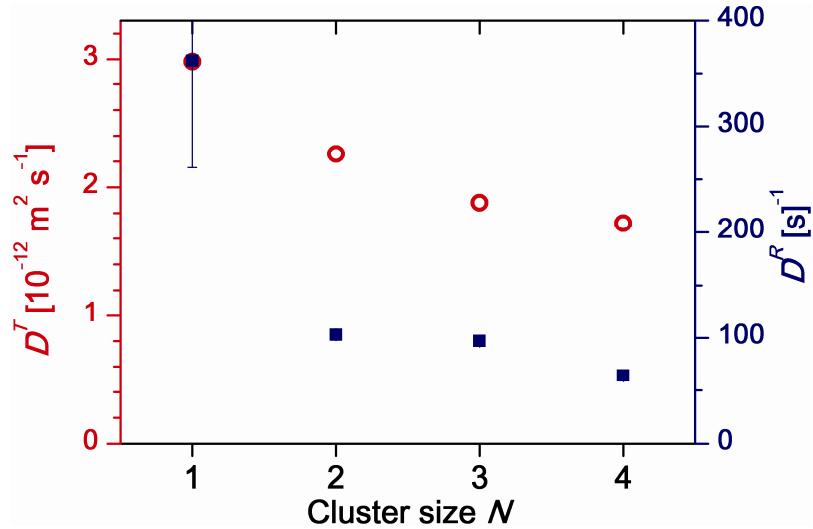

**Figure 3.** Translational diffusion coefficients $D^T$ (red circles) and rotational diffusion coefficients $D^R$ (blue squares) of the particle clusters as the function of the number of building blocks $N$ (single particles, $N = 1$; particle doublets, $N = 2$; triplets, $N = 3$; tetrahedrons, $N = 4$).

Figure 3 shows the translational and rotational diffusion coefficients of the different cluster species as derived from the $\Gamma$ versus $q^2$ plot according to equation (1) and (2). $D^T$ decreases smoothly with increasing size of the clusters. Translational diffusion is widely dominated by the volume of the assemblies, whereas the shape plays a minor role. Hence, the decrease of $D^T$ with the cluster size follows widely from the increase in the mean radius with the number of constituent particles of the clusters. However, the rotational characteristics differ from the translational properties. When going from the single particle to the particle doublet, a marked drop of $D^R$ is observed, whereas the particle triplet has almost the same rotational diffusion coefficient than the doublet (Figure 3). Similar observations were made for micron-sized clusters.[30]

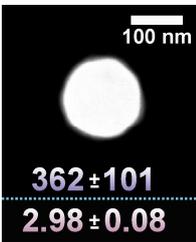

| Experiment | Model |
|---|---|
| $D^R$ [s$^{-1}$] | |
| $D^T$ [10$^{-12}$ m$^2$s$^{-1}$] | |
| 362 ± 101 | 359.0 ; $a$ = 80 nm, $p$ = 1 |
| 2.98 ± 0.08 | 3.06 |
| 103 ± 5 | 104.9$^\perp$ ; $a$ = 167 nm, $p$ = 0.5 |
| 2.26 ± 0.04 | 2.23 |
| 97 ± 2 | 92.5$^\perp$ ; $a$ = 76 nm, $p$ = 2 |
| 1.88 ± 0.04 | 1.95 |
| 64 ± 3 | 64.2 ; $a$ = 142 nm, $p$ = 1 |
| 1.72 ± 0.02 | 1.73 |

**Figure 4.** Translational ($D^T$) and rotational ($D^R_\perp$) diffusion coefficients as obtained by DLS and DDLS, respectively, together with the diffusion coefficients calculated for spheroids according to equations (3) – (7). The length of the axial semiaxis $a$ of the spheroids and the axial ratio $p$ are input estimates which can not be directly correlated with geometric parameters of the clusters. For the particle doublets and triplets as well as for the corresponding ellipsoids, $D_\perp^R$ perpendicular to the main axis of symmetry is measured and calculated, respectively. The SEM micrographs show the particle clusters oriented with their main body parallel to the plane of the figure (left column).

$D^R$ decreases again by 2/3, when going to the tetrahedron. Hence, it is evident that the decrease of $D^R$ does not exclusively correlate with the cluster volume because the rotational properties are strongly determined by the shape of the clusters as well. To gain further insight into the experimental data we use two different models which will be discussed in the following sections.

**Simple description of the dynamics on the basis of spheroids**

As mentioned above, the dynamics of the rigid clusters in solution encompass translational and rotational motion. These motions correlate directly with the overall size and shape of the clusters. The solvent is assumed to obey the low Reynolds number Navier-Stokes equation and incompressibility equations.[43] Hence, the hydrodynamic properties of a rigid object are contained in a diffusion matrix that provides a linear relationship between velocities and angular velocities to forces and torques acting on the body. In principle, the diffusion matrix can be calculated by solving the Navier-Stokes equation, but this is usually prohibitively difficult due to the complex shape of the objects. However, spherical bodies and spheroidal shapes are among the few shapes for which the flow equations of hydrodynamics can be solved exactly. For ellipsoids of revolution with two semiaxes of equal length the translational ($D_\parallel^T$, $D_\perp^T$) and rotational ($D_\parallel^R$, $D_\perp^R$) diffusion coefficients parallel and perpendicular to the main symmetry axis are given by:[43]

$$D_\parallel^T = \frac{k_B T}{8\pi\eta a}\frac{(2-p^2)G(p)-1}{1-p^2}, \qquad (3)$$

$$D_\perp^T = \frac{k_B T}{16\pi\eta a} \frac{(2-3p^2)G(p)+1}{1-p^2}, \tag{4}$$

$$D_\parallel^R = \frac{3k_B T}{16\pi\eta a^3 p^2} \frac{1-p^2 G(p)}{1-p^2}, \tag{5}$$

$$D_\perp^R = \frac{3k_B T}{16\pi\eta a^3} \frac{(2-p^2)G(p)-1}{1-p^4}, \tag{6}$$

where $G(p) = \log\left(\frac{1+\sqrt{1-p^2}}{p}\right)/\sqrt{1-p^2}$ for $p < 1$, and $G(p) = \arctan\left(\sqrt{p^2-1}\right)/\sqrt{p^2-1}$ for $p > 1$.

Here $a$ is the semiaxis along the axis of revolution, $b$ are the equatorial semiaxes, and $p = b/a$ is the axial ratio. In the case of prolate ellipsoids, the axial ratio $p$ is smaller than 1 since the axial semiaxis $a$ is longer than the equatorial semiaxes $b$. Conversely, $p > 1$ in oblate ellipsoids because the axial semiaxis $a$ is shorter than the equatorial semiaxes $b$. Finally, spheres have an axial ratio of 1 because all three semiaxes are equal in length. The orientation-averaged translational diffusion coefficient can be expressed as:

$$D^T = \frac{D_\parallel^T + 2D_\perp^T}{3} \tag{7}$$

$D^T$ can be measured by DLS and DDLS according to equations (1) and (2). The rotational diffusion around the axis of revolution characterized by $D_\parallel^R$ can be detected provided the ellipsoidal particles exhibit an optical anisotropy of sufficient magnitude. Otherwise only the rotational diffusion coefficient $D_\perp^R$ can be measured by DDLS due to the particle shape anisotropy.

As a first approximation, one may use equations (3) – (7) in order to model the diffusion coefficients of the clusters under consideration. Using the temperature $T$ = 298.15 K and the solvent viscosity $\eta$ = 0.000891 Nsm$^{-2}$ as input into equations (3) – (7) leads to the diffusion coefficients given in Figure 4. These calculated diffusion coefficients are similar to the experimental data shown in the left column of Figure 4. Hence, the dynamics of the clusters can be described in terms of the dynamics of a sphere for $N$ = 1 and 4, a prolate ellipsoid for $N$ = 2, and an oblate ellipsoid for $N$ = 3, where $N$ is the number of the building blocks. However, the length of the semiaxis $a$ and the axial ratio $p$ cannot be determined directly from the size and shape of the particle clusters in the case of $N$ = 2, 3, 4, but these parameters follow from modeling the experimental data with the help of equations (1) – (7). Thus, this model gives already a first description of the diffusion of the clusters. However, because its parameters cannot be derived directly from the geometry of the clusters, hydrodynamic models were probed that allow incorporating the true shape.

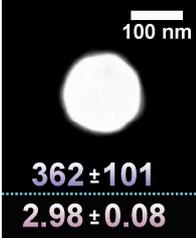

**Figure 5.** Comparison of the translational ($D^T$) and rotational ($D_{\parallel}^R$, $D_{\perp}^R$) diffusion coefficients as obtained by DLS and DDLS, respectively, together with the theoretical results using the shell model. For the particle doublets and triplets, the rotational diffusion coefficient $D_{\perp}^R$ perpendicular to the main

symmetry axis is measured. In the left column the particle clusters are oriented with their main body parallel to the plane of the figure.

**Modelling of the cluster dynamics based on hydrodynamic models**

In order to take into account the shape of the clusters correctly, we use the so-called shell model, in which the surface of the clusters is covered with a large number of non-overlapping spherical friction elements. This model is in widespread use and public-domain computer programs are available (see, e.g., Refs [28,44,51] and references therein). The hydrodynamic interaction between the beads is the crucial point in the numerical computation of the diffusion coefficients. Replacing a complex particle surface by a shell of very small spherical friction elements will give the correct diffusion coefficients, provided the diffusion matrix can be calculated numerically for a large number of small beads.

Figure 5 presents the calculated translational and rotational diffusion coefficients together with the experimental results. In the calculations the radius $R_s = 80$ nm of the constituent spheres and the center-to-center distance $L = 145$ nm between two spheres within a cluster was used. $L$ is taken directly from the field emission scanning electron microscopy (FESEM) images shown in Figure 1. It differs from the diameter of the building blocks because the clusters do not consist of touching constituent spheres. This is due to the assembly of the building blocks from toluene droplets. At the oil-water interface, the cross-linked polystyrene particles are partly swollen in toluene because toluene is a good solvent for polystyrene. Capillary forces created by the evaporation of the toluene pack the plasticized particles together in their final configuration. This leads to a partial deformation of the spheres and enhances the contact area among the constituent spheres within the cluster.

From the comparison of the experimental data with the calculated results, it is apparent that the shell model is an appropriate theoretical tool for studying the dynamics of these systems (Figure 5). In addition, we used a bead model[28,44] in which the particle clusters are represented directly by big spheres of radius $R_s$. Using the same model parameters as for the shell model leads to translational and rotational diffusion coefficients which differ by less than 3 % from the results of the shell model calculations. Hence, the bead model confirms the results obtained with the shell model.

We note that the experimental results cannot be explained assuming slip boundary conditions instead of conventional stick boundary conditions [11] which led to the theoretical data shown in Figures 4 and 5. For example, assuming slip boundary conditions would lead to an increase of the translational diffusion coefficient of a sphere by the factor of 1.5 as compared to the calculated values for $D^T$ in Figures 4 and 5, whereas the rotational motion of a sphere does not displace any fluid which implies a diverging rotational diffusion coefficient. Both results do not agree with the experimental data. Hence, the assumption of slip in contrast to stick boundary conditions is not appropriate for modeling the diffusion of the clusters. Slip boundary conditions work better for small objects of molecular dimensions which have lower orientation times than the particle clusters (Figure 2C).[11,52]

Modeling on the basis of the shell model was used to confirm the kind of rotation which is monitored in the DDLS experiment because diffusion coefficients for the rotation around all specific axes can be calculated and compared to the experimental result. A sphere does not possess specific axes of rotation. However, particle doublets and triplets have two specific axes of rotation (Figure 5). $D_{\parallel}^R$ characterizes the rotation around the main symmetry axis, *i.e.* the axis that connects the centers of the building blocks in the case of the doublet, whereas the main symmetry of the triplet is the C3 axis perpendicular to the plane of the constituents. $D_{\perp}^R$ is related to the rotation around the axis perpendicular to the main symmetry axis of the objects. In both cases the rotational diffusion coefficient $D_{\perp}^R$ is measured by DDLS due to the shape anisotropy of the particles, while $D_{\parallel}^R$ cannot be detected because of the rather small anisotropy of the spherical building blocks (Figure 5).

As discussed in the previous section, the diffusion of a tetrahedron resembles those of a spherical object because of its low shape anisotropy. For this reason, the rotation of the tetrahedron cannot be assigned to a specific axis.

To summarize this point, the shell model gave an excellent prediction of the diffusion coefficients of the particle clusters because the true shape as derived from the FESEM micrographs could be taken into account. For this reason, the predictions obtained from this model agree well with the experimental results (Figure 5).

**CONCLUSIONS**.

A common method in studying the dynamics of particles is DLS. DDLS further broadens the scope of this technique because it can be used to study both translational and rotational properties of small particles which underlie Brownian motion. Monodisperse spherical particles with diameters in the order of 100 nm in the range can be combined to different submicron-sized clusters with well-defined shape. For this reason, and because the translational and rotational properties of submicron-sized clusters are decoupled, colloidal clusters present ideal model systems to study the diffusion of complex particles with DDLS. Unlike diffusion studies by microscopy, DDLS monitors the true diffusion properties and does not underlie wall effects. A simple description of the diffusion of clusters made up from a small number of constituents is obtained in terms of the diffusion of rotational ellipsoids. The major disadvantage is that there is no direct relation between the geometric parameters of the clusters and the rotational ellipsoids. This gap between experiment and theory can be overcome by sophisticated hydrodynamic models such as the shell model which were developed in recent years. These models allow a precise prediction of the diffusion coefficients based on the shape of the objects. Moreover, they are very useful tools to interpret experimental data sets. Hence, the present study of the diffusion of submicron-sized particle clusters contributes to the fundamental understanding of the dynamics of particles with complex shape. Hence, it is intended for general use because the dynamics of complex particles are relevant to many practical problems occurring both in nature and industrial processes.

**METHODS**.

**Chemicals.** The chemicals used were purchased either from Sigma-Aldrich or Merck. Styrene was purified by washing with 10 wt% NaOH solution, drying over $CaCl_2$, and vacuum distillation. All other chemicals were of analytical grade and used as received.

**Cluster preparation.** Amino-modified polystyrene particles were used to build the clusters. These particles were prepared by emulsion polymerization of styrene with divinylbenzene (5 mol% relative to styrene) as the cross-linker, aminoethylmethacrylate hydrocloride (AEMH, 3 mol% relative to styrene) as the comonomer, cetyltrimethylammonium bromide (CTAB) as the emulsifier, and $\alpha,\alpha'$-azodiisobtyramidine dihydrochloride (V-50) as the initiator. The polymerization was carried out at 80 °C under a nitrogen atmosphere and continuous stirring at 320 rpm for 6 h. Purification of the latex particles was accomplished by exhaustive ultrafiltration against water. The size and the size distribution of the spherical particles were determined by dynamic light scattering (DLS), transmission electron microscopy, and disk centrifugation. The particles have an average diameter of 154 nm and can be regarded as monodisperse because their polydispersity index given as the the weight-average diameter divided by the number-average diameter is 1.004. The zeta potential of the particles bearing amino groups on their surface is +66 mV ± 5 mV.

The particle clusters were prepared along the lines given in Ref. [38]. This experimental approach was based on the agglomeration of the particles which were adsorbed onto the surface of narrow-dispersed emulsion droplets. Briefly, the particles were transferred from water into toluene. 3 ml of the 4.5 wt% suspension were added to a 0.5 wt% aqueous solution of surfactant (Pluronic F-68). A narrow-dispersed oil-in-water emulsion was obtained through powerful ultrasonication using a high-shear homogenizer. The self-assembly of the particles was induced by evaporation of the toluene using a rotary evaporator.

Separation of the suspension into fractions of clusters consisting of the same number of building blocks was accomplished through density gradient centrifugation. The density gradient was prepared by using a standard gradient maker with equal volumes of a 9 wt% and a 20 wt% aqueous glycerol solution.

Glycerol was used because it can be easily removed during the subsequent dialysis of the cluster fractions against water.

**Methods.** Field emission scanning electron microscopy (FESEM) was performed using a Zeiss LEO 1530 Gemini microscope equipped with a field emission cathode. Electrophoretic mobilities ($u$) of the particles were measured on a Malvern Zetasizer Nano ZS in conjunction with a Malvern MPT2 Autotitrator and converted into zeta potentials ($\zeta$) *via* the Smoluchowski equation ($\zeta = u\eta/\varepsilon_0\varepsilon$, where $\eta$ denotes the viscosity and $\varepsilon_0\varepsilon$ the permittivity of the suspension).

DLS and DDLS measurements were performed at 25 °C on a light scattering ALV/DLS/DLS-5000 compact goniometer system (Peters) equipped with a He-Ne laser (wavelength 632.8 nm), an ALV-6010/160 External Multiple Tau Digital Correlator (ALV), and a thermostat (Rotilabo). Cluster agglomerates and dust were removed through centrifugation (Beckman Coulter Allegra 64R) at 3000 rpm for 20 min. Prior to the measurement, the supernatant was filtered through 0.45 µm PET syringe filters (membraPure Membrex 25) into dust-free quartz glass cuvettes (Hellma). The volume fractions of the cluster suspensions were $10^{-5}$ to $10^{-6}$. The samples were placed in a *cis*-decaline index matching bath because *cis*-decaline does not change the polarization plane of the laser light. For each sample, three runs of 180 s (DLS) or 180 to 900 s (DDLS) were performed at scattering angles of 20 to 90° (DLS) or 20 to 60° (DDLS). The scattered light passed through a Glan Thomson polarizer (B. Halle) with an extinction ratio better than $10^{-5}$. CONTIN analysis of the intensity autocorrelation functions was used to calculate the relaxation frequencies.[47]

**ACKNOWLEDGMENT** The authors gratefully acknowledge financial support from the Deutsche Forschungsgemeinschaft (DFG) within SFB 840. M.H. thanks the Elite Network of Bavaria (ENB) within the graduate program "Macromolecular Science". A.W. is grateful to the Fonds der Chemischen Industrie (FCI), and the Dr. Otto Röhm Gedächtnisstiftung.

**Supporting Information Available**. Plots of the relaxation rates $\Gamma$ as obtained by DLS and DDLS for single particles, doublets, and triplets. Plot showing the contribution of the depolarized signal to the total scattering intensity for the different species. This material is available free of charge *via* the internet at http://pubs.acs.org.

SYNOPSIS TOC

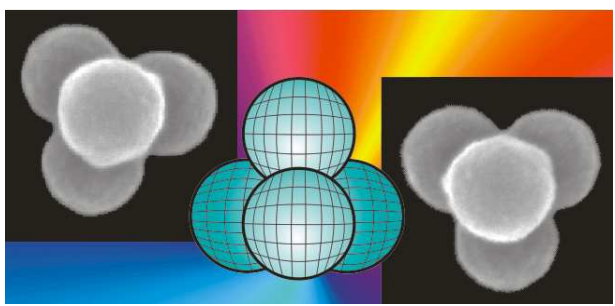